# Green InGaN/GaN LEDs: High luminance and blue shift

Anis Daami*[a], François Olivier[a], Ludovic Dupré[a], Christophe Licitra[a], Franck Henry[a], François Templier[a], Stéphanie Le Calvez[a]

[a]Université Grenoble Alpes, CEA-LETI, Minatec Campus, III-V Lab, Grenoble, France

## ABSTRACT

We report in this paper electro-optical results on InGaN/GaN based green micro light-emitting diodes (µLEDs). Current-light-voltage measurements reveal that the external quantum efficiency (EQE) behavior versus charge injection does not follow the ABC model prediction. Light-emission homogeneity investigation, carried out by photoluminescence mapping, shows that the Quantum Confinement Starck Effect (QCSE) is less significant at the edges of µLEDs. Electroluminescence shows a subsequent color green-to-blue deviation at high carrier injection levels. The extracted spectra at different current injection levels tend to show the appearance of discrete wavelength emissions. These observations may enhance the hypothesis that higher-energy excited-levels in InGaN quantum wells may also contribute to the blue shift, solely attributed to QCSE lessening under intense electric field magnitudes. We hereby present first results dealing with green µLEDs electro-optical performances with regards to their size.

## 1. INTRODUCTION

Micro light-emitting diodes (µLEDs) based on InGaN/GaN quantum well structures have seen a significant amount of progress in the last two decades[1, 2, 3]. New emerging applications such as augmented/mixed or virtual reality are looking forward using micro-displays (µ-displays) based on these still progressing technologies. However, a majority of studies in literature are mostly dealing with blue emitting µLEDs, with fewer investigation papers on green light emitting ones. Actually, the high-content indium incorporation in InGaN alloys, to adjust the emission wavelength to green color, is not an easy process track to deal with.

Nonetheless, green µLEDs are well positioned to cover high luminance needing components. Indeed, the eye sensitivity is at its maximum in this range of emitted wavelengths. Furthermore, for specific applications such as see-through glasses, a high brightness level of 5000 cd/m² or more is a mandatory keystone to achieve. Nevertheless, quantum efficiency of µLEDs is yet a challenging subject when dealing with high luminance levels. Besides, µLED size dependence has been thoroughly studied and reported in literature, but mostly for blue color emitting ones[4, 5]. A lack of comprehension of size effect is then still to be filled for these green light-emitting diodes dedicated to specific µ-displays applications.

This paper deals with an electro-optical study on different sized green µLEDs. First, quantum efficiency performance is analyzed versus µLED size. In a second part, we focus respectively, on µLED emission homogeneity and spectral response behaviors versus electrical injection.

## 2. ELECTRO-OPTICAL INVESTIGATION

### 2.1 Current-Voltage characteristics

Current density versus voltage characteristics of the different measured devices are shown on figure 1. At first glance, the current density seems independent of the µLED size on a large bias range. Larger devices (500µm and 200µm) show a slightly lower current density at high voltage values that is attributed to a probable deviation of the series resistance from an ideal geometrical law. We have shown in a recent study[6] that a series resistance variation can have a major impact on the µLED current at high voltage. On the other hand, at a bias value lower than the µLED threshold voltage, the perceived current density difference is mainly related to the limitation of our apparatus sensitivity. Despite these slight observed differences, we demonstrate once again the robustness of our µLED fabrication process[7].

*anis.daami@cea.fr

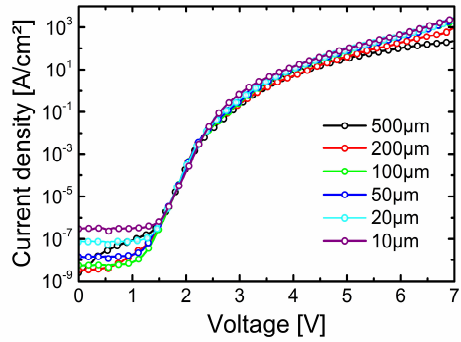

Figure 1. Current density versus voltage characteristics of different square µLEDs. Current density is constant on a large bias range whatever the µLED size is. The legend denotes the side width of each µLED.

## 2.2 Optical power and quantum efficiency

As regards to optical properties of our devices, figure 2(a) shows the measured optical output power versus applied voltage of the different µLEDs. The corresponding external quantum efficiency (EQE) of each device are plotted versus current density on figure 2(b). In contrary to what is reported[4] on our blue LEDs, the optical threshold voltage ($V_{TO}$) of green µLEDs does not seem to vary much in a certain range of geometry (width ≥ 50µm). The extracted value of $V_{TO}$ in this range of geometry is around 2.1 ± 0.05V. For smaller µLEDs, $V_{TO}$ begins to shift, attaining 2.5 V for a 10µm µLED.
Moreover, the optical output power of large devices, shows two distinct regimes. At low voltage, a first rapid increase is observed. Then, this fast upturn slows down, showing a 'kink' like effect. In a second time, the power rises again with bias until a saturation regime due to droop is attained. The echo of these two regimes is evident on the EQE curves of large devices. Indeed a first increase is perceived, then the EQE reaches a plateau (EQE ~ 2%), corresponding to the optical power output 'kink' region. Afterwards, the EQE restarts to increase reaching a maximum $EQE_{max}$ = 5%, before sinking down due to droop effect.
It is worth pointing out that the first rapid regime seems to disappear when the size of the µLED shrinks. This is evident on EQE curves where the plateau regime tends to vanish for small geometries. Besides, the EQE threshold current density moves to higher values for the small devices (width ≤ 20 um). We hereafter, emit the hypothesis that the fading of the rapid optical output regime and the shift of the optical threshold voltage $V_{TO}$ are correlated.

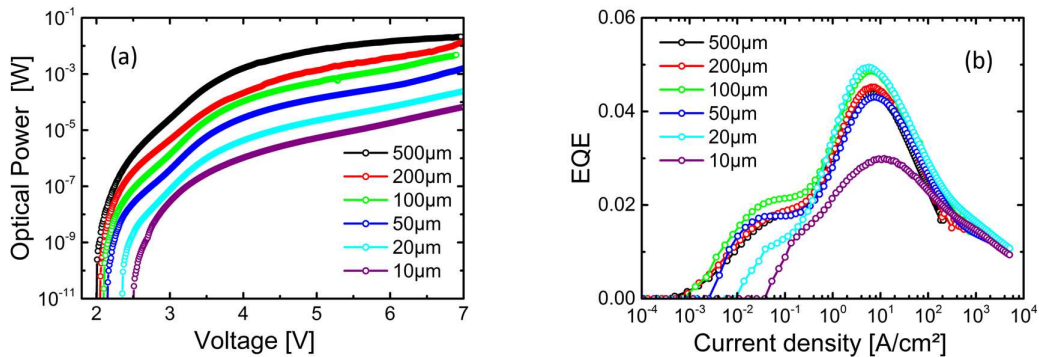

Figure 2. (a) Optical power versus voltage and (b) external quantum efficiency versus current density characteristics of different square µLEDs. Two different optical regimes are depicted on large scale µLEDs. The 1st regime tends to disappear for smaller µLEDs. It is worth noting that the 1st regime vanishing is correlated to the shift of the optical emission threshold also observed on smaller geometry µLEDs.

## 2.3 Light emission homogeneity

Many studies in literature point out the importance of the µLED periphery on its optical efficiency[8, 9]. Downsizing the device geometry has usually come out with lower optical efficacies. One crucial adopted reason is the perimeter taking over the surface, when the µLED size diminishes. Numerous investigations in this direction, have shown the increase of SRH non-radiative recombination. This is primarily related to defects and a bad passivation of µLED periphery after etching steps. We have carried out photoluminescence (PL) mapping on a 3 by 3 array of green µLEDs (7x7 µm²) to check the quality of our etching process and light emission homogeneity. Figure 3 shows respectively, the obtained PL intensity and wavelength mappings. Oddly, we observe a higher signal at the edges of the µLEDs. Moreover the emitted wavelength is to some extent lower at these boundaries. This observation, presumably points out a less predominant QCSE at the periphery of our devices. Indeed, we suppose a strong relaxation of the lattice constraint at the edges of the µLED. This constraint reduction gives rise to a reduced spontaneous polarization in the quantum wells. Hence, the locally emitted light is shifted to a higher energy. Moreover, the reduced band curvature would allow a better recovery of hole and electron wave-functions, explaining the stronger emission intensity. A similar observation has been reported[10] by Xie et al. using the cathodoluminescence technique.

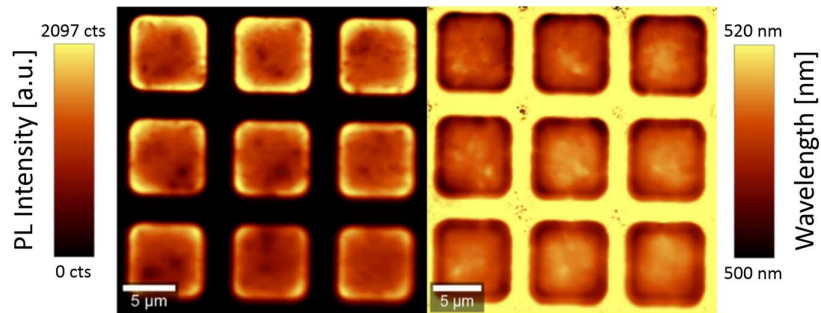

Figure 3. Photoluminescence intensity and wavelength mappings on a 3x3 7µm squared µLEDs array. A higher intensity correlated to a shorter wavelength emission is observed on the edges of µLEDs. We suppose a probable lattice relaxation at the edges of µLEDs, reducing the QCSE effect behind this observation.

Another important observation shown in figure 4 reveals that light emission homogeneity is bias dependent on larger scale µLEDs (100x100 µm² LED example shown). Indeed we can see that at low voltage range, the emission is rather speckled on the surface of the device. This spattered pattern is mainly related to a non-homogeneous indium incorporation in quantum wells. Consequently, this indium spread results in a local optical threshold variation through the device surface itself. Higher content indium spots will have a lower optical threshold voltage, hence emitting at lower bias values. It is regularly reported in literature that high content indium amalgamation with GaN is a harsh path to consider[11, 12]. Another reason to this light speckle pattern may also come from a local variation of contact resistivity. This electrical deviation can be related to the indium spread, stated above. It can also be due to a non-homogeneous P doping, locally degrading the contact/semiconductor interface[13, 14].

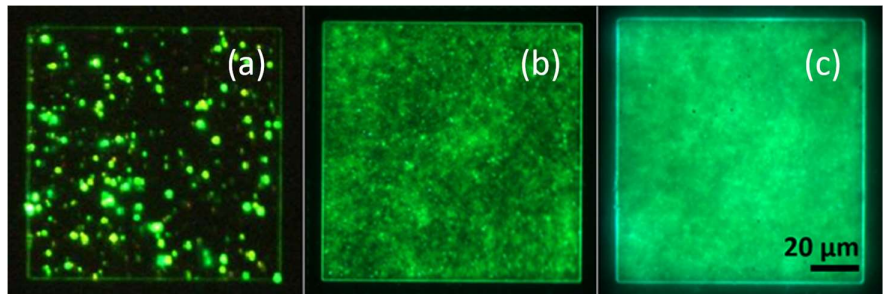

Figure 4. Optical photos of light emission homogeneity on a 100x100 µm² green LED at different bias values: (a) V = 2.2V, (b) V = 3.5V and (c) V = 4.5V. Note the high luminance spread at low voltage that disappears at higher bias values.

Regarding smaller area µLEDs, usually dedicated to micro-displays, this singularity has to be banned. Indeed luminance homogeneity is one important key factor of micro-displays. This observed spread in luminance tends to disappear when injection bias is driven upwards. Nevertheless, this voltage increase induces the appearance of another oddity. Actually, the emitted-light wavelength shifts towards lower values. More details on this shift are discussed in the next section.

## 3. THE BLUE SHIFT DILEMMA

### 3.1 Electroluminescence and color shift

We carried out electroluminescence measurements on small sized unitary µLEDs (6µm diameter) in order to limit the surface-spread light-emission observed at low injection. Optical photos taken at different applied voltages, presented in figure 5, perceptibly show an intense wavelength shift of the emitted light when injection bias is increased. We clearly see a green to blue transformation of the emitted color.

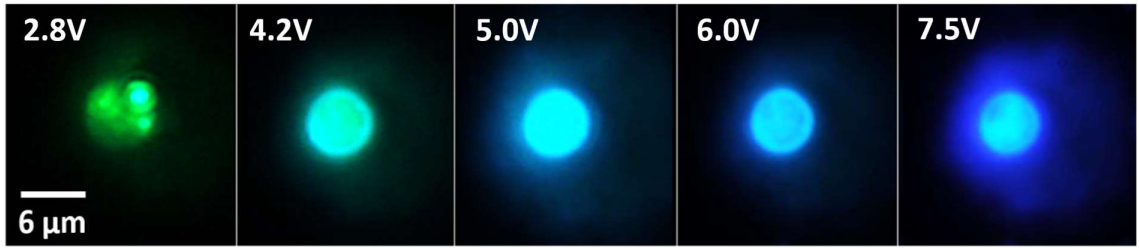

Figure 5. Optical photos of light emission on a 6µm diameter 'green' µLED at different bias values. The emitted wavelength is very dependent on applied voltage across the µLED. Green color turns to blue when increasing bias from 2.8V to 7.5V.

This important evolution has been quantified, and the color point plotted on a CIE chromaticity 1931 diagram, displayed in figure 6. Each measured point corresponds to an increase of 0.1V of the µLED bias, varying from 2.5V to 7.5V. There is no doubt that the emitted wavelength covers a large range on the CIE diagram varying from green ($\lambda$=540nm) to an almost true blue ($\lambda$=475nm).

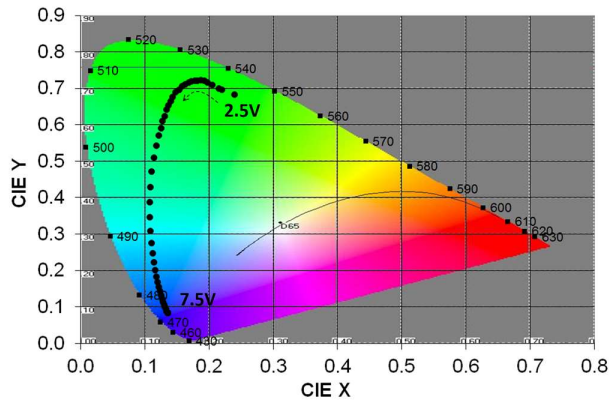

Figure 6. Color point evolution of the emitted light from a green µLED plotted on CIE chromaticity 1931 diagram. Bias varies from 2.5V to7.5V. Each point corresponds to 100mV increase in voltage. The measured wavelength varies from 540nm (green) to 475nm (blue).

The wavelength blue shift is also observed in blue emitting µLEDs but at a lesser extent. It is usually attributed to the fading, under high voltage, of the Quantum Confined Starck Effect (QCSE). This later is explained by the default presence of a local electric field in the quantum wells. Its magnitude is directly depending on both the spontaneous and piezo-electric polarizations existing in InGaN/GaN light emitting devices[15]. In the case of green µLEDs, the high indium content induces a huge lattice mismatch, at the barrier/quantum well interface between InGaN and GaN materials. Subsequently, a high charge density appears at the quantum well interfaces. Consequently, a high magnitude piezo-electric field, related to these pseudo charge-sheets, heavily twists the energy bands in the quantum well and the QCSE is more pronounced. This energy band bending induces a red shift in optical transitions between confined energy levels in quantum wells. When the injection voltage level is increased through the µLED, the piezo-electric field is slowly screened and the energy bands are less warped. The red shift linked to QCSE is then recovered and slowly vanishes turning into a blue shift. Nevertheless, blue shift elucidations are still debated in literature.

### 3.2 Quantum well excited-levels filling hypothesis

We present on figure 7, normalized intensity electroluminescence spectra at four different voltage values, recorded on the 6µm µLED. At low bias, close to $V_{TO}$, one wavelength peak centered at 520 nm is observable (red curve). This peak corresponds to the mean indium content incorporated in the InGaN quantum well. When the voltage increases, we observe the formation of new emission peaks alongside the first one. At 3.5V, beside the 520 nm peak, we perceive the appearance of a higher intensity one, centered at 508 nm, and a slight hump at 480 nm (green curve). After increasing the bias by 2V, (blue curve) the 480 nm intensity takes over the two precedent higher wavelength peaks, and a hump at 460 nm begins to appear. This later shows the same luminescence intensity as the 480 nm peak when voltage is raised to 7.5V (purple curve). At this bias level, both precedent peaks (520 nm and 508 nm) appear as small bulges compared to the high energy peaks (460 nm and 480 nm). The discrete wavelength peaks tend to demonstrate that high energy excited levels inside quantum wells are progressively populated at high injection levels. This permits the appearance of higher energy, hence shorter wavelength optical transitions. This wavelength and peak intensity ballet versus injection bias is a plausible elucidation of the green to blue shift in our µLEDs.

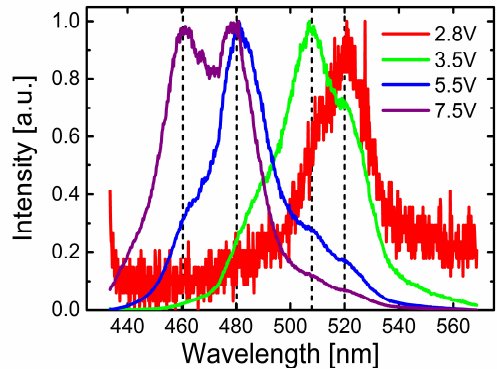

Figure 7. Normalized electroluminescence spectra versus bias carried out on a 6 µm diameter µLED. In contrary to what is usually believed, the blue shift in InGaN/GaN µLEDs is not solely due to the QCSE. The appearance of discrete emission peaks with variable intensities depending on injection bias level tends to corroborate the filling of high excited levels in the quantum wells explaining the color changing behavior in green µLEDs.

## 4. CONCLUSION

InGaN/GaN based green µLEDs are less debated in literature compared to blue emitting ones. One main reason is the difficulty to incorporate high indium content in InGaN alloys. Yet, they are good candidates for simple see-through applications as the eye sensitivity to green is at its maximum. We have exposed through an electro-optical thorough analysis of different sized green µLEDs that their quantum efficiency behavior is far from being easily described by an easy ABC model. In addition, they show a less effective QCSE at their edges. The light emission homogeneity study also reveals an indium content dispersion inside quantum wells. The electroluminescence measurements divulges a drastic green to blue color shift. Beside QCSE reduction in µLEDs under bias, we suspect the filling of high energy levels in quantum wells as a likely explanation of this shift. Finally, this paper aims to a better understanding of InGaN/GaN based green µLEDs features, by a very first investigation of size effect on their electro-optical behavior.

## ACKNOWLEDGEMENTS

The authors acknowledge partial fundings from the European Union's Horizon 2020 VOSTARS research and innovation programme under grant agreement No 731974 and H2020 HILICO European project (H2020- JTI-CS2-2016-CFP04-SYS-01-03, Grant No. 755497).